\documentstyle[12pt,aps,prl,epsfig]{revtex}

\begin{document}
\title{Quantum Arnol'd diffusion in a rippled waveguide}
\author{V.~Ya.~Demikhovskii$^{1}$, F.~M.~Izrailev$^{2}$ and A.~I.~Malyshev$^{1}$}
\address{
$^{1}$Nizhny Novgorod State University, 603950, Nizhny Novgorod,
Gagarin ave. 23, Russia\\
$^{2}$Instituto de F\'isica, Universidad Aut\'onoma de Puebla,
Apdo. Postal J-48, Puebla 72570, Mexico\\}

\date{\today}
\maketitle
\begin{abstract}
We study the quantum Arnol'd diffusion for a particle moving in a
quasi-1D waveguide bounded by a periodically rippled surface, in
the presence of the time-periodic electric field. It was found
that in a deep semiclassical region the diffusion-like motion
occurs for a particle in the region corresponding to a stochastic
layer surrounding the coupling resonance. The rate of the quantum
diffusion turns out to be less than the corresponding classical
one, thus indicating the influence of quantum coherent effects.
Another result is that even in the case when such a diffusion is
possible, it terminates in time due to the mechanism similar to
that of the dynamical localization. The quantum Arnol'd diffusion
represents a new type of quantum dynamics, and may be
experimentally observed in measurements of a conductivity of
low-dimensional mesoscopic structures.
\end{abstract}

\newpage

As is well known, one of the mechanisms of the dynamical chaos in
the Hamiltonian systems is due to the interaction between
nonlinear resonances~\cite{C59}. When the interaction is strong,
this leads to the so-called global chaos which is characterized by
a chaotic region spanned over the whole phase space of a system,
although large isolated islands of stability may persist. For a
weak interaction, the chaotic motion occurs only in the vicinity
of separatrices of the resonances, in accordance with the
Kolmogorov-Arnol'd-Moser (KAM) theory (see, for example, Ref.~\cite{1}).
In the case of two degrees of freedom ($N = 2$), the
passage of a trajectory from one stochastic region to another is
blocked by KAM surfaces.

The situation changes drastically in many-dimensional ($N > 2$)
systems for which the KAM surfaces no longer separate stochastic
regions surrounding different resonances, and chaotic layers of
the destroyed separatrices form a stochastic web that can cover
the whole phase space. Thus, if the trajectory starts inside the
stochastic web, it can diffuse throughout the phase space. This
weak diffusion {\it along} stochastic webs was predicted by
Arnol'd in 1964~\cite{2}, and since that time it is known as a
very peculiar phenomenon, however, universal for many-dimensional
nonlinear Hamiltonian systems (see, for example, review~\cite{3}
and references therein).

Recently, much attention has been paid to the chaotic dynamics of
a particle in a rippled channel (see, for example,
Refs.~\cite{4,5,6,Lent,7}). The main interest was in the
quantum-classical correspondence for the conditions of a strong
chaos. Specifically, in Ref.~\cite{4} the transport properties of
the channel in a ballistic regime were under study. Energy band
structure, the structure of eigenfunctions and density of states
have been calculated in~\cite{5,6}. The quantum states in the
channel with rough boundaries, as well as the phenomena of quantum
localization have been analyzed in~\cite{7}. The influence of an
external magnetic field for narrow channels was investigated
in~\cite{Lent}. These studies may have a direct relevance to the
experiments with periodically modulated conducting channels. In
this connection one can mention the investigation ~\cite{Kou} of
transport properties of a mesoscopic structure (sequence of
quantum dots) with the periodic potential formed by metallic
gates.

In contrast with the previous studies, below we address the regime
of a weak quantum chaos which occurs along the nonlinear
resonances in the presence of an external periodic electric field.
Our goal is to study the properties of the quantum Arnol'd
diffusion which may be observed experimentally. The approach we
use is based on the theory developed in Ref.\cite{9} by making use
of a simple model of two coupling nonlinear oscillators, one of
which is driven by two-frequency external field.

We study the Arnol'd diffusion in a periodic quasi-one dimensional
waveguide with the upper profile given in dimensionless variables
by the function $y=d+a \cos{x}$. Here $x$ and $y$ are the
longitudinal and transverse coordinates, $d$ is the average width,
and $a$ is the ripple amplitude. The low profile is assumed to be
flat, $y=0$. The nonlinear resonances arise due to the coupling
between two degrees of freedom, with the following resonance
conditions,
\begin{equation}
  \eta =T_x/T_y =\omega_y /\omega_x
  \label{res_cond}
\end{equation}
Here $T_y$ is the period of a transverse oscillation inside the
channel, $T_x$ is the time of flight of a particle over one period
of the waveguide, $\omega_x$ and $\omega_y$ are the corresponding
frequencies, and $\eta$ is the rational number.

\begin{figure}[htb]
\begin{center}
\epsfig{file=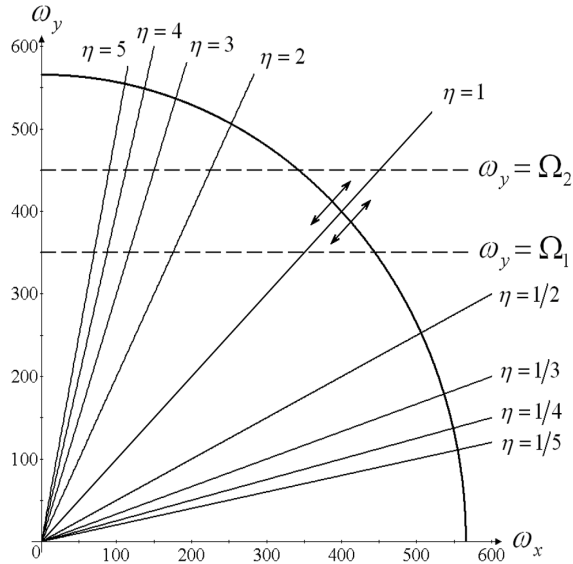,width=3.0in,height=3.0in}
\vspace{0.8cm}
\caption{Some of the coupling resonances for different values of
$\eta$, with the isoenergetic curve $E=160000$ in the frequency
plane. The locations of resonances are shown by dashed lines.}
\end{center}
\end{figure}

The mechanism of the classical Arnol'd diffusion in this system is
illustrated in Fig.~1. Here some of the resonance lines for
different values of $\eta$ are shown on the  $\omega_x$,
$\omega_y$-plane. Also, the curve of a constant energy is shown
determined by the equation
\begin{equation}
  \label{res_omega}
  \omega_x^2 + \left(\frac{\omega_y d}{\pi}\right)^2 = 2mE.
\end{equation}
Here $E$ and $m$ are the dimensionless kinetic energy and particle
mass, respectively, which we set to unity in what follows. The
neighboring coupling resonances are isolated one from another by
the KAM-surfaces, therefore, for a weak perturbation the
transition between their stochastic layers is forbidden. Such a
transition could occur in the case of the resonance overlap, i.e.
in the case of the global chaos only. In the absence of an
external field, the passage of a trajectory along any stochastic
layer (this direction is shown at Fig.~1 by two arrows) is also
impossible because of the energy conservation. However, the
external time-periodic field removes the latter restriction, and a
slow diffusion along stochastic layers becomes possible.

An external electric field with the corresponding potential
$V(y,t)=-f_0y(\cos \Omega_1 t + \cos \Omega_2 t)$ gives rise to
two main resonances with the frequencies $\omega_y = \Omega_1$ and
$\omega_y=\Omega_2$. Their locations are shown in Fig.~1 by the
dashed lines. In order to calculate the diffusion rate, we
consider a part of the Arnol'd stochastic web created by three
resonances, namely, by the coupling resonance and two guiding
resonances with frequencies $\Omega_1$ and $\Omega_2$. Therefore,
we chose the initial conditions inside the stochastic layer of the
coupling resonance. To avoid the overlapping of the resonances, we
assume the relation $f_0/a=1000$ is fulfilled.

For the further analysis it is convenient to pass to the
curvilinear coordinates $x_i^{'}$ in which both boundaries are
flat~\cite{8}. The covariant coordinate representation of the
Schr\"odinger equation has the following form,
\begin{equation}
   \label{schr}
  -\frac{1}{2 \sqrt {g}} \frac {\partial}{\partial
   x_i^{'}} \sqrt {g} g_{ij} \frac {\partial \psi}{\partial x_j^{'}}
   =E \psi
\end{equation}
where $g_{ik}$ is the metric tensor, $g\equiv \det (g_{ij})$. Here
we use the units in which the Plank's constant and effective mass
are equal to unity. As a result, the new coordinates are
\begin{equation}
  \label{transf} x'=x, \,\,\,\,\,\,\,\,\,\,\, y'=\frac{y}{1+\epsilon\cos x}
\end{equation}
where $\epsilon=a/d$. In these coordinates the boundary conditions
are $\psi(x',0)=\psi(x',d)=0$ and the metric tensor is
\begin{equation}
g_{ij}= \left(
             \matrix{
             1 \,\,\,& \frac{\epsilon x' \sin x'}{1+\epsilon \cos x'} \cr
             \frac{\epsilon x' \sin x'}{1+\epsilon \cos x'}
             \,\,\,&
             \frac{1+\epsilon ^2 x'^2 \sin ^2 x'}{(1+\epsilon \cos x')^2}
                    }
         \right),
\label{tensor}
\end{equation}
with the orthonormality condition,
\begin{equation}
   \label{ortho}
   \int \psi_i^{\star} \psi_j\sqrt{g}dS'=\delta_{ij}.
\end{equation}
If the ripple amplitude $a$ is small compared to the channel width
$d$, than keeping only the first-order terms in $\epsilon$ in the
Schr\"odinger equation~(\ref{schr}), we obtain the following
Hamiltonian~\cite{8},
\begin{equation}
\label{ham} \hat H = \hat H_0(x,y) + \hat U(x,y),
\end{equation}
where
\begin{equation}
\label{H0} \hat H_0 = - \frac{1}{2} \left(
\frac{\partial^2}{\partial x^2} + \frac{\partial^2}{\partial y^2}
\right)
\end{equation}
and
\begin{eqnarray}
\label{H00} \hat U = \frac {\epsilon}{2} \left( 2\cos x
\frac{\partial^2}{\partial y^2} -2 y \sin x
\frac{\partial^2}{\partial x \partial y} - y \cos x
\frac{\partial}{\partial y} - \frac{1}{2} \cos x - \sin x
\frac{\partial }{\partial x} \right) .
\end{eqnarray}
Here and below we omitted primes in coordinates $x'$ and $y'$.

Since the Hamiltonian is periodic in the longitudinal coordinate
$x$, the eigenstates are Bloch states. This allows us to write the
solution of the Schr\"odinger equation in the form $\psi(x,y)=\exp
(ikx) \Phi _k (x,y)$ where $\Phi_k (x+2\pi, y) = \Phi _k (x,y)$.
For an infinite periodic channel the Bloch wave vector $k$ has
continuous values, in particulary, $-1/2 \leq k \leq 1/2$ in the
first Brillouin zone.

By expanding $\Phi_k (x,y)$  in the double Fourier series the
eigenstates can be written as
\begin{equation} \psi^k (x,y) =
  e^{ikx} \sum_{n,m} c^k_{nm} \psi^0_{nm}(x,y)
  \label{psi}
\end{equation}
where
\begin{equation}
   \label{psi0}
   \psi^0_{nm}(x,y)=\sqrt {\frac{2}{Ld}}
   e^{inx}\sin \left( \frac {\pi my}{d}\right)
\end{equation}
are the eigenstates of the unperturbed Hamiltonian $\hat H_0(x,y)$
with the corresponding eigenvalues
\begin{equation}
E^0_{nm} = \frac{1}{2} \left( (n+k)^2 + \frac{\pi ^2 m^2}{d^2}
\right )
\end{equation}
for the considered case with $L \gg 2\pi$. Now we proceed to solve
the system of algebraic equations for the coefficients $c^k_{nm}$,
\begin{equation}
E(k)c^k_{nm} = \frac{1}{2} \left( (n+k)^2 + \frac{\pi ^2 m^2}{d^2}
\right ) c^k_{nm} + \sum _{n',m'} U_{(k+n)m, (k+n')m'} c^k_{nm}.
\label{system}
\end{equation}
Here the matrix elements are
\begin{equation}
  \label{matel}
  \matrix{
  U_{(k+n)m, (k+n')m'}=\int \left( \psi^0_{(k+n'),m'}
  \right)^{\star} \hat U(x,y) \psi^0_{(k+n),m} dx dy =  \cr
  - \frac{a}{2d} \left[ \frac{\pi^2 m^2}{d^2} \left(
  \delta_{n',n+1} + \delta_{n',n-1} \right) \delta_{m,m'} +
  \frac{\left(-1\right)^{m+m'}mm'}{m^2-m'^{2}}
   \left( \left( 1+2(k+n)\right)
  \delta_{n',n+1} + \left( 1-2(k+n)\right) \delta_{n',n-1} \right)
  \right].
  }
\end{equation}
Following Refs.~\cite{9}, we analyze the dynamics in the vicinity
of the main coupling resonance $\eta=1$ which is determined by the
condition $\omega_{n_0}=\omega_{m_0}$ where $\omega_{n_0} =
E_{n_0+1}(k)- E_{n_0}(k)= k + n_0 +1/2$ and
$\omega_{m_0}=E_{m_0+1}- E_{m_0}=\pi^2(2m_0+1)/2d^2$. In a deep
semiclassical region where $n_0 \gg 1$ and $m_0 \gg 1$, one can
write $\omega_{n_0} \approx n_0$ and $\omega_{m_0} \approx \pi^2
m_0/d^2$. It should be noted that the similar resonance condition
can be satisfied for negative $n_0$ also in the case when
$-n_0\approx\pi^2 m_0/d^2$, which corresponds to the particles
moving in the opposite direction. Below we use the fact that when
$|n_0|\gg 1$ the two resonances (two sets of states),
corresponding to $n_0>0$ and $n_0<0$ are not coupled for all $k$
apart from the center and edges of the Brillouin zone. This fact
is due to the anti-unitary symmetry of the Hamiltonian for all
values of $k$, apart from the indicated above. The properties of
the energy spectra and eigenstates for specific values $k=\pm 1/2$
and $k=0$ will be discussed elsewhere.

In the vicinity of the resonance it is convenient to introduce new
indexes $r=n-n_0$ and $p=r+(m-m_0)$. Then, instead of the system
(\ref{system}) we obtain,
\begin{equation}
   \label{systemlast}
   E(k) c^k_{rp} = p \omega_{m_0}c^k_{rp} + \frac{1}{2} \left( r^2 +
   \frac{\pi^2}{d^2}\left( p-r \right)^2 \right) c^k_{rp} +
   \sum_{r',p'} U_{k+n_0-r, p-r+m_0, k+n_0 -r', p'-r'+m_0} c^k_{r'p'}.
\end{equation}
Here we count energy $E(k)$ from the level $E^0_{n_0m_0}(k)$.

Numerical calculation of the solution of these equation gave the
following results. First, as in Ref.\cite{9}, the energy spectrum
consists of a number of Mathieu-like groups corresponding to the
coupling resonance. These groups are separated one from another by
the energy  $\omega_{n_0}$. The structure of energy spectrum in
each group is typical for a quantum nonlinear resonance. Inside
the resonance, the lowest levels are practically equidistant, the
accumulation point corresponds to the classical separatrix and all
states are non-degenerate. The states above the separatrix are
quasi-degenerate due to to the rotation in opposite directions.

In accordance with the spectrum structure it is convenient to
characterize the states at coupling resonance by two indexes:
group number $q$ and $s$ --- level number inside the group.
Correspondingly, the energy of each group can be written as follows,
\begin{equation}
  \label{energy}
  E_{q,s}(k) =\omega_{n_0}(k) q + E_{q,s}^{M}(k),
\end{equation}
where $E_{q,s}^{M}$ is the Mathieu-like spectrum for one group.
The indexes $q$ and $s$ correspond to fast and slow variables
characterizing the motion inside the classical coupling resonance.

Let us consider now the dynamics of a charged particle in the
rippled channel in the presence of the time dependent electric
field described by the potential $V(y,t)= - f_0 y (\cos \Omega_1 t
+ \cos \Omega_2 t)$. We assume that the frequencies $\Omega_1$ and
$ \Omega_2 $  are chosen to fulfill the condition $\omega_{n_0} =
\left( \Omega_1 + \Omega_2 \right)/2$ in order to provide equal
driving forces for a particle inside the stochastic layer of the
separatrix under consideration. Specifically, we take,
$\omega_{n_0}=400, \,\, \Omega_1 = 350, \,\, \Omega_2 =450$,
therefore, the period $T$ of the perturbation is, $T=7\cdot
2\pi/\Omega_1 = 9 \cdot 2\pi/ \Omega_2 \approx 0.126$.

Since the total Hamiltonian is periodic in time, one can write the
solution of the non-stationary Schr\"odinger equation as
$\psi(x,y,t) = \exp \left( -i\epsilon_Q t\right) u_Q (x,y,t)$,
where $u_Q (x,y,t)$ is the quasienergy (QE) function and
$\epsilon_q$ is the quasienergy. As is known, the QE functions are
the eigenfunctions of the evolution operator $\hat U (T)$ of the
system for one period of the perturbation. The procedure to
determine this operator was described in details in Ref.~\cite{9}.
The matrix elements $U_{q,s,q',s'}(T)$ of the evolution operator
can be calculated by means of the numerical solution of the
non-stationary Schr\"odinger equation. Then, the evolution matrix
$U_{q,s,q',s'}(NT)$ for $N$ periods can be easily obtained.

Our goal is to analyze the dynamics of a particle placed inside
the separatrix under the condition that the coupling and two
driving resonances do not overlap. The evolution of any initial
state can be computed using the evolution matrix as follows,
\begin{equation}
  \label{CC}
  C_{q,s}(NT)=\sum _{q',s'} U_{q,s,q',s'}(NT)C_{q',s'}(0).
\end{equation}

\begin{figure}[htb]
\begin{center}
\epsfig{file=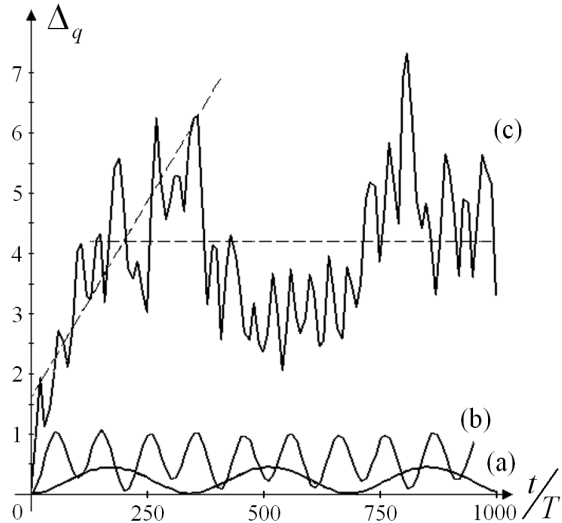,width=3.0in,height=2.8in}
\vspace{0.8cm}
\caption{Time dependence of the variance $\Delta_q$ for different
initial states at coupling resonance for the group with $q = 0$:
(a) the lowest level with $s = 0$, (b) the above-separatrix level
with $s = 45$, (c) the near separatrix level with $s = 22$. Here
$a = 0.01 $ and $f_0 = 10$. }
\end{center}
\end{figure}

In Fig.~2 typical dependencies of the variance $\left(\Delta \bar
H \right)^2=\omega^2_{n_0}\Delta_q$ of the energy are shown versus
the time measured in the number $N$ of periods of the external
perturbation, for different initial conditions. The quantity $\Delta_q$
is defined as follows,
\begin{equation}
\label{deltaQ} \Delta_q= \sum_q (q-\bar q)^2\sum_s |C_{q,s}|^2,
\,\,\,\,\,\,\,\,\, \bar q = \sum_q q \sum_s |C_{q,s}|^2 .
\end{equation}

The data clearly demonstrate a different character of the
evolution of the system in dependence on the initial state.  For
the state taken from the center of the coupling resonance, as well
as above the separatrix, the variance oscillates in time, in
contrast with the state taken from inside the separatrix. In the
latter case, after a short time the variance of the energy
increases linearly in time, thus manifesting a diffusion-like
spread of the wave packet.

In order to characterize the speed of the diffusion, we have
calculated the classical diffusion coefficient $D_{cl}$ for the
coupling resonance $\eta=1$, in the comparison with the quantum
one, $D_q$, for different goffer amplitude and initial wave number
$k \neq 0; \pm 1/2$, see Fig.~3. It was found that the quantum
Arnol'd diffusion roughly corresponds to the classical one.
However, one can see a systematic deviation which indicates that
the quantum diffusion is weaker than the classical Arnol'd
diffusion. This deviation is due to the influence of coherence
quantum effects which can be very strong even in a deep
semiclassical region. As was shown in Ref.~\cite{10}, these
stabilizing quantum effects are enhanced for the motion inside
narrow stochastic layers surrounding the nonlinear resonances.
This effect is due a relatively small number $M_s$ of quantum
states belonging to the region of the separatrix, the fact which
is crucial in the study of the quantum-classical correspondence
for the systems with a weak chaos in the classical limit.

\begin{figure}[htb]
\begin{center}
\epsfig{file=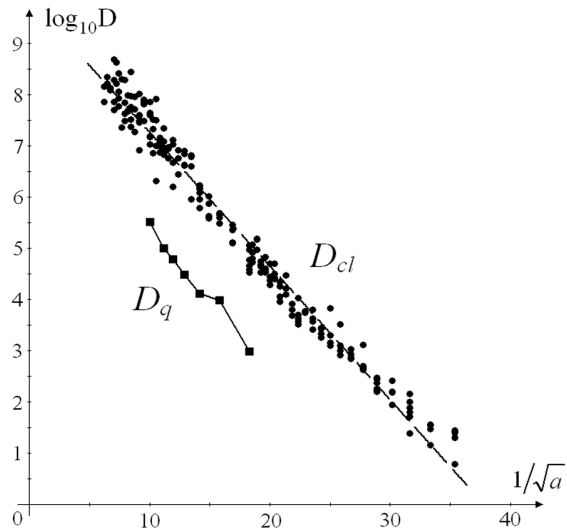,width=3.0in,height=2.8in}
\vspace{0.8cm}
\caption{Classical $D_{cl}$ versus quantum $D_q$ diffusion
coefficients in the dependence on the amplitude $a$ of the rippled
profile. }
\end{center}
\end{figure}

In this connection it is instructive to estimate the number of the
energy states that occupy the separatrix layer in our
calculations. We have found that for $a = 0.01$ of the amplitude
characterizing the upper profile of the waveguide, the number of
stationary states in the separatrix chaotic layer is about
$M_s\approx 10$, therefore, one can treat our results
semiclassically. Our additional calculation has shown that with
the decrease of the amplitude~$a$, the number~$M_s$ decreases, and
for $1/\sqrt a \approx 20$ this number is of the order one. For
this reason the last right point in Fig.~3 corresponds to the
situation when the chaotic motion along the coupling resonance is
completely suppressed by quantum effects. This effect is known as
the \lq\lq{}Shuryak border\rq\rq~\cite{10}) which establishes the
conditions for a complete suppression of classical chaos.

Since the diffusion along the coupling resonance is effectively
one-dimensional, one can expect an Anderson-like localization
which is known to occur in the presence of any weak disorder in
low-dimensional structures. Indeed, the variance of the QE
eigenstates of the evolution operator is finite in the $q$-space.
This means that eigenstates are localized, and the wave packet
dynamics in this direction has to reveal the saturation of the
diffusion. More specifically, one can expect that the linear
increase of the variance of the energy ceases after some
characteristic time.

In order to observe the dynamical localization in our model (along
the coupling resonance inside the separatrix layer), one needs to
analyze a long-time dynamics of wave packets. Our numerical study
for large times (see curve (c) in Fig.~2) have revealed that after
some time $t\sim 200T$, the diffusion-like evolution terminates
for all range of the amplitude~$a$. Specifically, on a large time
scale the variance starts to oscillate around a mean value. This
effect is of the same origin as the so-called dynamical
localization which was discovered in the kicked rotor
model~\cite{11,12}. One should note that the dynamical
localization is, in principle, different from the Anderson
localization, since the latter occurs for the models with random
potentials. In contrast, our model is the dynamical one (without
any randomness), and a kind of pseudo-randomness is due to the
quantum chaos mechanism.

In conclusion, we have studied the quantum Arnol'd diffusion for a
particle in the quasi-1D rippled waveguide under the time-periodic
electric field. This diffusion is known to occur in the
corresponding classical systems, below the threshold of an overlap
of nonlinear resonances resulting in a strong chaos. As is known,
in this case the chaotic motion is possible only for the
trajectories inside narrow stochastic layers surrounding the
nonlinear resonances. The classical Arnol'd diffusion is
exponentially weak, however, it leads to the unbounded motion
inside the resonance web created by the resonances of different
orders.

The problem under study was to find out whether the Arnol'd
diffusion is feasible in quantum systems describing the electron
motion in quasi-1D waveguides. Our results clearly demonstrate the
peculiarities of the quantum Arnol'd diffusion and establish the
conditions under which it can be observed experimentally. One of
the most important results is that the quantum diffusion is
typically weaker than the corresponding classical one. The
analysis has shown that the reason is the influence of quantum
coherent effects which are essentially strong for the motion in
the vicinity of the nonlinear resonances, in the case when these
resonances do not overlap. Moreover, we found that with a decrease
of the perturbation responsible for the creation of stochastic
layers, the quantum Arnol'd diffusion can be completely
suppressed, thus leading to the absence of any chaos in the
system.

Another important result is the observation of the dynamical
localization of the Arnol'd diffusion, which is manifested by the
termination of the diffusion on a large time scale. We have
briefly discussed the mechanism of this phenomenon, relating it
with the famous Anderson localization known to occur in
low-dimensional disordered structures. Thus, in addition to the
Shuryak border, the dynamical localization is a new mechanism
destroying the quantum Arnol'd diffusion.

Recently, the weak electron diffusion was observed experimentally
~\cite{From} in superlattices with stationary electric and
magnetic fields. At certain voltage, the unbounded electron motion
along a stochastic web changes the conductivity of the system and
results in a large increase of the current flow through the
superlattice. Similarly, one can expect that electron Arnol'd
diffusion inside resonance stochastic layers can increase the high
frequency conductivity of a rippled channel, and can be detected
experimentally.

This work was supported by the program
\lq\lq Development of scientific potential of high school\rq\rq{}
of Russian Ministry of Education and Science
and partially by the CONACYT (M\'exico) grant No~43730.
A.I.M. acknowledges the support of non-profit foundation
\lq\lq Dynasty\rq\rq.


\end{document}